\documentstyle[aps,prl,floats,twocolumn]{revtex}

\newcommand{\ep}{\epsilon}
\newcommand{\phb}{\bar{\phi}}
\newcommand{\lp}{{\cal L}_p}
\newcommand{\rs}{{\cal R}_s}

\title{Instabilities and disorder of the domain patterns in the systems
  with competing interactions} 
\author{C. B. Muratov}
\address{Department of Physics, Boston University, Boston, Massachusetts
  02215} \date{\today}

\draft

\begin{document}
\twocolumn[\hsize\textwidth\columnwidth\hsize\csname
@twocolumnfalse\endcsname

\maketitle

\begin{abstract}
  The dynamics of the domains is studied in a two-dimensional model of
  the microphase separation of diblock copolymers in the vicinity of the
  transition. A criterion for the validity of the mean field theory is
  derived. It is shown that at certain temperatures the ordered
  hexagonal pattern becomes unstable with respect to the two types of
  instabilities: the radially-nonsymmetric distortions of the domains
  and the repumping of the order parameter between the neighbors. Both
  these instabilities may lead to the transformation of the regular
  hexagonal pattern into a disordered pattern.
\end{abstract}


\bibliographystyle{prsty}

\vskip2pc]

Formation of complex patterns consisting of domains with sharp walls is
a beautiful example of self-organization in the systems both near and
far from thermal equilibrium
\cite{cross93,ko:book,niedernostheide,kapral,bates90,seul95}.  Recently,
it became clear that long-range competing interactions are responsible
for the formation of the domain patterns in the systems as diverse as
ferroelectrics and ferrofluids, garnet ferromagnets, Langmuir
monolayers, type I superconductors in the intermediate state, diblock
copolymers, reaction-diffusion systems with long-range inhibition (see
\cite{seul95} and references therein). In such systems the formation of
the uniform state favored by the local properties of the system is
precluded by the long-range interaction which does not favor that
uniform state. Thus, the system becomes separated into the domains of
the alternating values of the ``order parameter''. As a result of this
separation a lot of different equilibrium configurations, including
highly symmetric ones, are possible.

In this Letter we will study the time-dependent model of a system with
competing interactions in two dimensions. We will investigate the
stability of the stationary states and show that they undergo
instabilities which may change both the characteristic length scale and
the morphology of the domain patterns. We will also show that the
destabilization of the highly symmetric patterns typically leads to the
formation of the disordered patterns.

The microphase separating diblock copolymer melts are a typical example
of the systems with the long-range competing interactions. There the
macroscopic phase separation of the mutually incompatible monomers is
not allowed since the monomers are connected through the polymer chains.
Ohta and Kawasaki obtained the free energy for this system in the case
of the long polymer molecules \cite{ohta86}
\begin{eqnarray}
  \label{F}
  F = \int && d^d x \left( {(\nabla \phi)^2 \over 2} + {a \phi^2 \over
    2} + {b \phi^4 \over 4} \right. \nonumber \\ && \left. + {\alpha
    \over 2} \int d^d x' (\phi(x) - \phb) G(x - x') (\phi(x') - \phb)
\right),
\end{eqnarray}
where $\phi$ is the order parameter which is proportional to the
difference of the monomer concentrations, $a$ and $b$ are the
coefficients of the Landau-Ginzburg expansion, the last term in Eq.
(\ref{F}) is the long-range interaction characterized by the function
$G$ which reflects the connectivity of the chains, $\alpha \sim N^{-2}$
is the strength of this interaction and $N$ is the number of monomers in
a chain, the constant $\phb$ is determined by the block ratio. The
function $G$ satisfies
\begin{equation}
  \label{G}
  - \nabla^2 G(x - x') = \delta^{(d)} (x - x').
\end{equation}
The model given by Eqs. (\ref{F}) and (\ref{G}) in fact has a wider
applicability and can be used to describe the stationary states in
ceramic compounds with the long-range Coulombic interactions
\cite{chen93}, ferroelectric semiconductors \cite{mamin94},
high-temperature superconductors and degenerate magnetic semiconductors
\cite{nagaev95}, reaction-diffusion systems with the long-range
inhibitor \cite{ko:book,goldstein96}, and reaction-controlled spinodal
decomposition \cite{glotzer95}.

The formation of the domains in the system under consideration is due to
the competition between the nonlocal interaction and the surface tension
which is determined by the local terms in Eq. (\ref{F}). For the
equilibrium pattern the contributions of these two effects have to be
comparable. According to Eq. (\ref{F}), for the domain of size $R$ we
have
\begin{equation}
  \label{comp}
  \sigma R^{d-1} \sim N^{-2} \phi^2 R^{d+2},
\end{equation}
where $\sigma$ is the coefficient of the surface tension. Near the
critical point $\sigma = \sigma_0 |t|^{\nu (d - 1)}$, where $t = (T -
T_c)/T_c$, $T_c$ is the critical temperature, and $\nu$ is the critical
exponent of the Ising model \cite{widom65}. Because of the long-range
character of the nonlocal interaction, in estimating its contribution
one can ignore the fluctuations of $\phi$ and put $\phi = \phi_0
|t|^\beta$ in Eq. (\ref{comp}), where $\beta$ is the respective critical
exponent of the Ising model. Then, according to Eq.  (\ref{comp}), the
characteristic size of the equilibrium domain near the critical point
will be
\begin{equation}
  \label{R}
  R \sim N^{2/3} |t|^{\nu (d - 1) - 2 \beta \over 3}.
\end{equation}
Substituting this expression into Eq. (\ref{comp}), we obtain that the
free energy of a single domain $\Delta F \gg 1$, if
\begin{equation}
  \label{mft}
  N \gg |t|^{2 \beta - \nu (d + 2) \over 2}.
\end{equation}
If this condition is satisfied, we can use the mean-field approximation
to describe the domain patterns in the vicinity of the phase transition.

In order to study the domain patterns in the mean-field limit let us
introduce the new variables
\begin{equation}
  \label{new}
  \phi' = {\phi \over \phi_0 |t|^{\beta}}, ~~ x' = {x \over \xi}, ~~
  \ep^2 = {2 \sqrt{2} \alpha \phi_0^2 \xi_0^3 \over 3 \sigma_0 } |t|^{2
    \beta - \nu (d + 2)},
\end{equation}
where $\xi = \xi_0 |t|^{-\nu}$ is the correlation length, and write the
free energy functional (up to a constant factor and dropping the primes)
\begin{eqnarray}
  F = \int && d^d x \left( {(\nabla \phi)^2 \over 2} - {\phi^2 \over 2}
  + {\phi^4 \over 4} \right. \nonumber \\ && \left. + {\ep^2 \over 2}
  \int d^d x' (\phi(x) - \phb) G(x - x') (\phi(x') - \phb) \right).
  \label{Fn}
\end{eqnarray}
One should think of Eq.  (\ref{Fn}) as the mean-field representation of
the domain interactions. When the condition of Eq.  (\ref{mft}) is
satisfied, we have $\ep \ll 1$, what implies the {\em strong segregation
  limit} \cite{ohta86}. Notice that in these units $R \sim \ep^{-2/3}$.

The evolution of the oder parameter $\phi$ near the phase transition in
the mean-field limit can be described by the time-dependent
Landau-Ginzburg equation obtained from Eq. (\ref{Fn}) \cite{fizkiny}.
This equation can be written as a pair of reaction-diffusion equations
of the activator-inhibitor type in the limit of the fast inhibitor:
\begin{eqnarray}
  {\partial \phi \over \partial t} = \nabla^2 \phi + \phi - \phi^3 -
  \psi, \label{act} \\ 0 = \ep^{-2} \nabla^2 \psi + \phi - \phb,
  \label{inh}
\end{eqnarray}
where $\phi$ plays the role of the activator and $\psi$ plays the role
of the inhibitor \cite{ko:book}, and the time scale has been absorbed
into the definition of $t$. In the limit $\ep \rightarrow 0$ these
equations are equivalent to the interfacial dynamics problem
\cite{goldstein96,m:pre96}. We would like to emphasize that in this
limit Eqs. (\ref{act}) and (\ref{inh}) are applicable to the systems
with the conserved order parameter. Indeed, the equations of the
interfacial dynamics mentioned above can be alternatively derived from
the free energy given by Eq. (\ref{Fn}) if one assumes the overdamped
dynamics of the interface \cite{goldstein96}.  The latter is justified
for the microphase separation in the diblock copolymer melts, where the
hydrodynamic effects are important.

The domain patterns that form in this kind of systems have been recently
studied in a great detail in connection with the patterns forming in
highly nonequilibrium systems
\cite{m:pre96,mo1:pre96,mo2:pre96,m2:pre96}.  Muratov and Osipov showed
that in the case of Eqs. (\ref{act}) and (\ref{inh}) only the patterns
whose characteristic length scale (in these units) is of order
$\ep^{-2/3}$ can be linearly stable \cite{m:pre96,mo1:pre96,mo2:pre96}.
Thus, if the condition in Eq.  (\ref{mft}) is satisfied, on this length
scale the mean-field approximation remains valid for the time-dependent
theory as well.

When $|\phb| < 1$ and $\ep \ll 1$, Eqs. (\ref{act}) and (\ref{inh})
admit the solutions in the form of the stationary multidomain patterns,
including the hexagonal patterns of circular domains of radius $\rs$ and
the period $\lp$ in two dimensions
\cite{ko:book,ohta86,m:pre96,mo1:pre96,mo2:pre96,m2:pre96}. It is clear
that the period of these patterns (or, actually, the characteristic
interdomain distance) must lie within $l \ll \lp \lesssim L$, where $l
\sim 1$ and $L \sim \ep^{-1}$ are the characteristic length scales of
the variation of $\phi$ and $\psi$, respectively \cite{ko:book}. For
$|\phb|$ not very close to 1 the radius of the domains is comparable to
$\lp$. Since the characteristic size of the domains is of order
$\ep^{-2/3}$, the period of the pattern must also be $\lp \sim
\ep^{-2/3} \ll L$, so the value of $\phi$ is close to $+1$ inside the
domains and close to $-1$ outside.

Let us now study the stability of the hexagonal pattern. It is clear
that the results obtained for this pattern should be qualitatively the
same for an arbitrary multidomain pattern.  In this case, according to
Eq. (\ref{inh}), the relationship between the period and the radius of
the domains is
\begin{equation}
  \label{rs}
  \rs = 3^{1/4} \lp \left( { 1 + \phb \over 4 \pi } \right)^{1/2},
\end{equation}
so the ratio $\rs / \lp$ can be conveniently used as a parameter instead
of $\phb$. According to the general asymptotic theory of instabilities
of domain patterns in reaction-diffusion systems \cite{mo1:pre96}, the
dangerous fluctuations of $\phi$ are localized in the domain walls and
represent the small distortions of the domains. The damping decrement
$\gamma$ of such fluctuations is determined by the eigenvalues of a
certain operator.  Since the hexagonal pattern we are interested in is
periodic, we can partially diagonalize this operator by considering the
fluctuations modulated by the wave vector $k$ which lies in the
Brillouin zone and obtain for the system under consideration
\cite{mo1:pre96} 
\begin{equation}
  \label{disp}
  \left( -\gamma + {m^2 \over \rs^2} + \lambda_0 \right)
    \delta_{mm'} = - 3 \sqrt{2} ~\ep^2 \rs R_{mm'} (k),
\end{equation}
where $m$ and $m'$ are integers corresponding to the modes of the
azimuthal distortions of the domains, $\lambda_0$ is a constant
independent of $m$ and $k$, $\delta_{mm'}$ is the Kronecker delta, and
$R_{mm'}(k)$ is a certain matrix with the indices $m$ and $m'$. It is
possible to show (the details of the derivations will be given
elsewhere) that the constant $\lambda_0$ is given by
\begin{equation}
  \label{l0}
  \lambda_0 = - {1 \over \rs^2} - {3 \ep^2 \rs \over \sqrt{2}} \left( 1
  - {2 \pi \rs^2 \over \sqrt{3} \lp^2} \right).
\end{equation}

\noindent
Since the characteristic length scale of the pattern is much smaller
than $L$, the Laplacian dominates in the equation for the Green's
function involved in the calculation of $R_{mm'}(k)$ (see Ref.
\cite{mo1:pre96}), so this Green's function may be assumed to satisfy
Eq.  (\ref{G}).  With this fact in mind it is convenient to write the
expression for $R_{mm'}(k)$ in the Fourier space 
\begin{eqnarray}
  \label{Rmm}
  && R_{mm'}(k)= \nonumber \\ &&~~~{2 \pi \over v} \sum_{k'} { e^{i (m'
      - m) (\vartheta_{k + k'} + \frac{\pi}{2}) } \over | k + k'|^2 }
  J_m (|k + k'| \rs) J_{m'} (|k + k'| \rs), \nonumber \\ 
\end{eqnarray}
where $k'$ runs over the reciprocal lattice, $J_m$ are the Bessel
functions, $\vartheta_{k + k'}$ is the angle between the vector $k + k'$
and the $x$-axis, and $v = \sqrt{3} \lp^2/2$ is the volume of the
elementary cell. Notice that for $\rs \ll \lp$ the off-diagonal elements
of $R_{mm'}(k)$ become negligibly small, so for $|m| \ge 2$ we recover
the instabilities of the localized domains --- autosolitons
\cite{mo1:pre96}. As $\rs$ gets bigger, the mixing between the vectors
corresponding to the different values of $m$ occurs. This mixing,
however, is not very strong, so one can still label the eigenvectors of
Eq. (\ref{disp}) with $m$.

As was shown qualitatively by Kerner and Osipov, for the most dangerous
fluctuations the wave vector $k$ will lie close to the edge of the
Brillouin zone \cite{ko:book}. There are two basic types of the
fluctuations we need to consider: the fluctuations with $m = 0$ which
lead to repumping of the order parameter between the neighboring domains
(Fig.  \ref{instab}[a]) and the fluctuations with $m = 2$ which lead to
the asymmetric distortions of the domains (Fig.  \ref{instab}[b])
\cite{ko:book}. The analysis of Eq. (\ref{disp}) shows that the most
dangerous fluctuations with $m = 0$ have $k = \frac{1}{3} (b_1 - b_2)$,
where $b_1$ and $b_2$
\noindent
are the reciprocal lattice vectors which make a $120^{\rm o}$ angle,
while the most dangerous fluctuations with $m = 2$ have $k = \frac{1}{2}
(b_1 + b_2)$.  The instability $\gamma < 0$ occurs with respect to
repumping when $\lp < {\cal L}_{p0}$ or with respect to the asymmetric
distortion when $\lp > {\cal L}_{p2}$, where ${\cal L}_{p0,2}$ depend on
$\ep$ and $\rs/ \lp$. The resulting stability diagram is presented in
Fig. \ref{phase}. Figure \ref{phase} also shows the period of the
equilibrium hexagonal pattern obtained by Ohta and Kawasaki
\cite{ohta86}. One can see that the equilibrium hexagonal pattern is
stable for all values of $\rs$ (except, possibly, for $\rs / \lp$ close
to 0.5 where the assumption about the circular shape of the domains
ceases to be valid). In two dimensions the equilibrium hexagonal
pattern corresponds to the global minimum of the free energy if $\rs /
\lp < 0.31$, whereas for $0.31 < \rs/\lp < 0.37$ (the second condition
means that $\phb < 0$, i.e., positive domains in the negative
background) the global minimum corresponds to the lamellar pattern
\cite{ohta86}. Figure \ref{phase}, however, does not show the transition
from the hexagonal to the lamellar pattern, so in fact the equilibrium
hexagonal pattern is always metastable.

Recall that the values of $\ep$ and $\phb$ strongly depend on
temperature near the critical point. Suppose that the equilibrium
hexagonal pattern formed as a result of the slow quench of the system
below $T_c$. If now the temperature is abruptly raised, the domains may
become unstable with respect to the asymmetric distortions.  To study
the kinetics of this process we solved Eqs.  (\ref{act}) and (\ref{inh})
numerically with the initial condition in the form of the hexagonal
pattern plus small noise. This process is shown in Fig.  \ref{sim}(a).
At the end of the simulation the system reaches an asymptotic state. One
can see that the hexagonal pattern transform into a highly disordered
metastable pattern which consists of the domains of complex shapes.
Notice that this effect was observed experimentally in different systems
with competing interactions \cite{cape71,heckl86}.

If the system temperature is quickly lowered, the
pattern may destabilize with respect to repumping.  The kinetics of this
process is shown in Fig.  \ref{sim}(b).  As the destabilization
progresses, some of the domains start to ``eat'' their neighbors, what
results in either increase of the interdomain distance or fusion of some
of the neighbors.  A lot of the domains shrink and eventually disappear.
All these processes create a lot of disorder.  Eventually the distance
between the domains becomes large enough, so that the repumping is no
longer realized, and the resulting pattern orders somewhat at late
stages.  However, the asymptotic pattern in the end of the simulation is
still highly disordered. Notice that the repumping effect is 
observed in experiments and numerical simulations of systems with
competing interactions \cite{coulon93,sagui94}. 

In summary, we showed that the stationary multidomain patterns in the
systems with long-range competing interactions may undergo
instabilities affecting both the period and the shape of the domains
which typically result in the formation of highly disordered
metastable patterns.

This work was partially supported by the Material Theory Program, DMR
NSF.


\begin{thebibliography}{10}

\bibitem{cross93}
M. Cross and P.~C. Hohenberg, Rev. Mod. Phys. {\bf 65},  851  (1993).

\bibitem{ko:book}
B.~S. Kerner and V.~V. Osipov, {\em Autosolitons: a New Approach to Problems of
  Self-Organization and Turbulence} (Kluwer, Dordrecht, 1994).

\bibitem{niedernostheide}
{\em Nonlinear dynamics and pattern formation in semiconductors and devices},
  edited by F.~J. Niedernostheide (Springer, Berlin, 1994).

\bibitem{kapral}
{\em Chemical waves and patterns}, edited by R. Kapral and K. Showalter
  (Kluwer, Dordrecht, 1995).

\bibitem{bates90}
F.~S. Bates and G.~H. Fredrickson, Annu. Rev. Phys. Chem. {\bf 41},  525
  (1990).

\bibitem{seul95}
M. Seul and D. Andelman, Science {\bf 267},  476  (1995).

\bibitem{ohta86}
T. Ohta and K. Kawasaki, Macromolecules {\bf 19},  2621  (1986).

\bibitem{chen93}
L.~Q. Chen and A.~G. Khachaturyan, Phys. Rev. Lett. {\bf 70},  1477  (1993).

\bibitem{mamin94} R.~F. Mamin, Pis'ma Zh. Eksp. Teor. Fiz. {\bf 60}, 51
  (1994) [JETP Lett. {\bf 60}, 52 (1994)].

\bibitem{nagaev95}
E.~L. Nagaev, Usp. Fiz. Nauk {\bf 165}, 529 (1995) [Phys. Uspekhi {\bf
  38}, 497 (1995)]. 

\bibitem{goldstein96}
R.~E. Goldstein, D.~J. Muraki, and D.~M. Petrich, Phys. Rev. E {\bf 53},  3933
  (1996).

\bibitem{glotzer95}
S. Glotzer, E.~A. Di~Marzio, and M. Muthukumar, Phys. Rev. Lett. {\bf 74},
  2034  (1995).

\bibitem{widom65} B.~Widom, J. Chem. Phys. {\bf 43}, 3898 (1965).

\bibitem{fizkiny} E.~M. Lifshits and L.~P. Pitaevskii, {\em Physical
    kinetics} (Pergamon Press, Oxford, 1981).

\bibitem{m:pre96}
C.~B. Muratov, Phys. Rev. E {\bf 54},  3369  (1996).

\bibitem{mo1:pre96}
C.~B. Muratov and V.~V. Osipov, Phys. Rev. E {\bf 53},  3101  (1996).

\bibitem{mo2:pre96}
C.~B. Muratov and V.~V. Osipov, Phys. Rev. E {\bf 54},  4860  (1996).

\bibitem{m2:pre96}
C.~B. Muratov,  Phys. Rev. E (to be published). 

\bibitem{cape71}
J.~A. Cape and G.~W. Lehman, J. Appl. Phys. {\bf 42},  5732  (1971).

\bibitem{heckl86} W.~M. Heckl and H. M\"{o}hwald, Ber. Bunsenges. Phys.
  Chem. {\bf 90}, 1159 (1986).

\bibitem{coulon93} G. Coulon, B. Collin, and D. Chatenay, J. de Physique
  II {\bf 3}, 697 (1993).

\bibitem{sagui94}
C. Sagui and R. Desai, Phys. Rev. E {\bf 49},  2225  (1994).

\end{thebibliography}

\begin{figure}[htbp]
  \vspace{0.5cm}
  \caption{Two major types of instabilities of the hexagonal pattern.}
  \label{instab}
\end{figure}

\begin{figure}[htbp]
  \caption{Stability diagram for the hexagonal pattern. The dashed
    line corresponds to the equilibrium pattern.}
  \label{phase}
\end{figure}

\begin{figure}[htbp]
 \vspace{0.5cm}
  \caption{(a) Destabilization of the hexagonal pattern with respect to the
    asymmetric distortions. Distributions of $\phi$ at different times
    for $\ep = 0.05$ and $\phb = 0$. The system is $200 \times 230$. (b)
    Destabilization of the hexagonal pattern with respect to repumping.
    Distributions of $\phi$ at different times for $\ep = 0.025$ and
    $\phb = -0.2$. The system is $200 \times 230$.}
  \label{sim}
\end{figure}

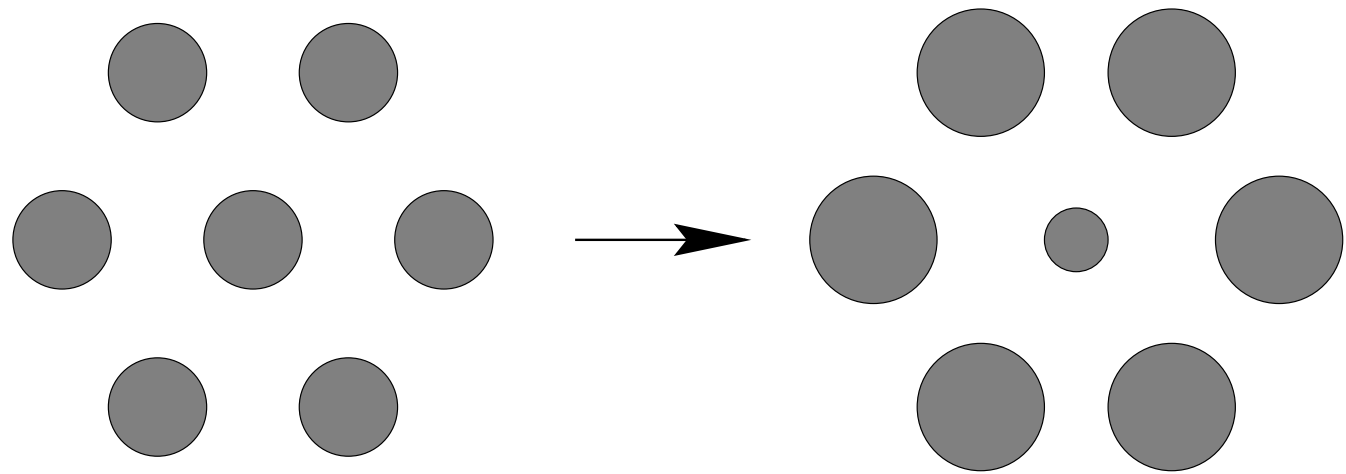
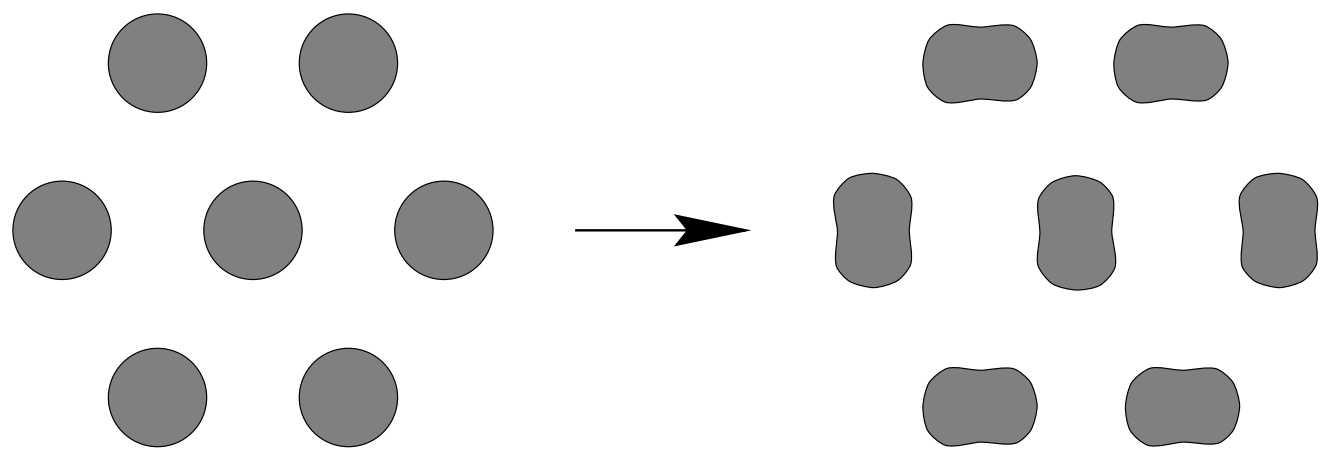
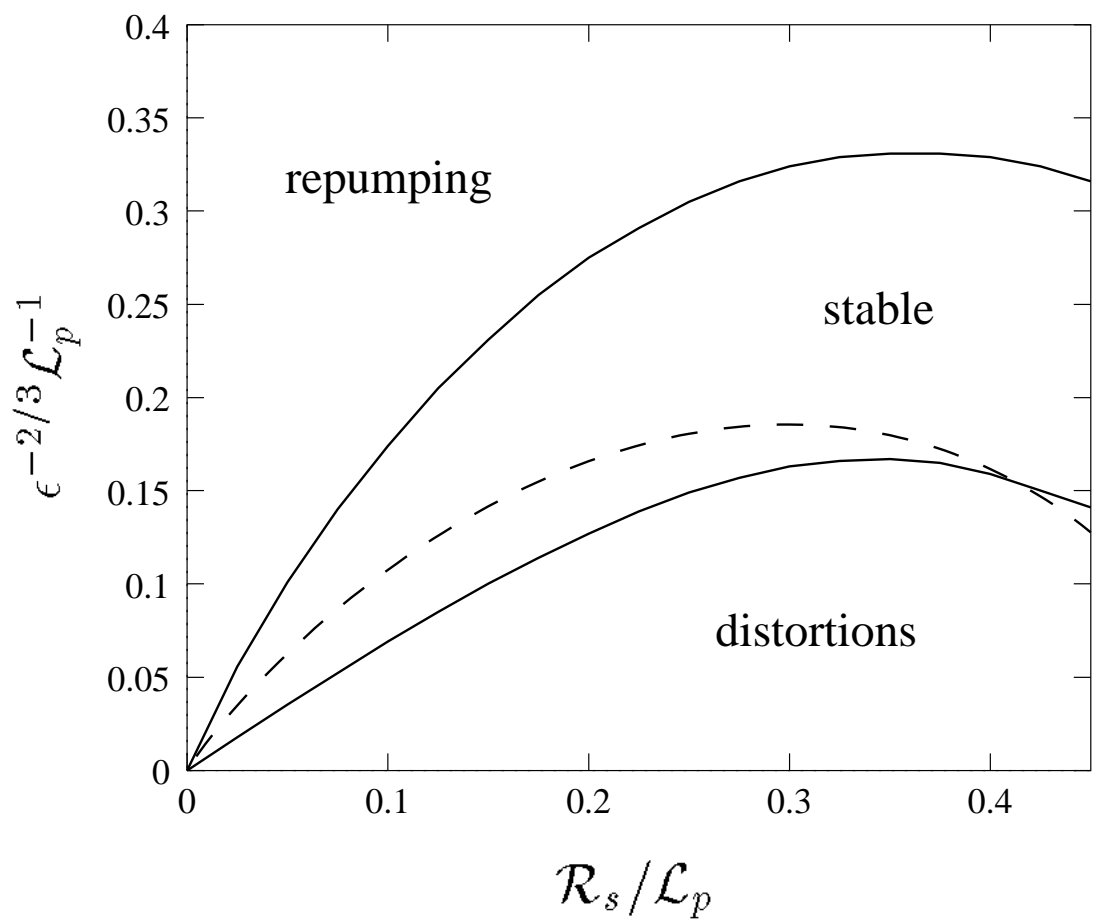
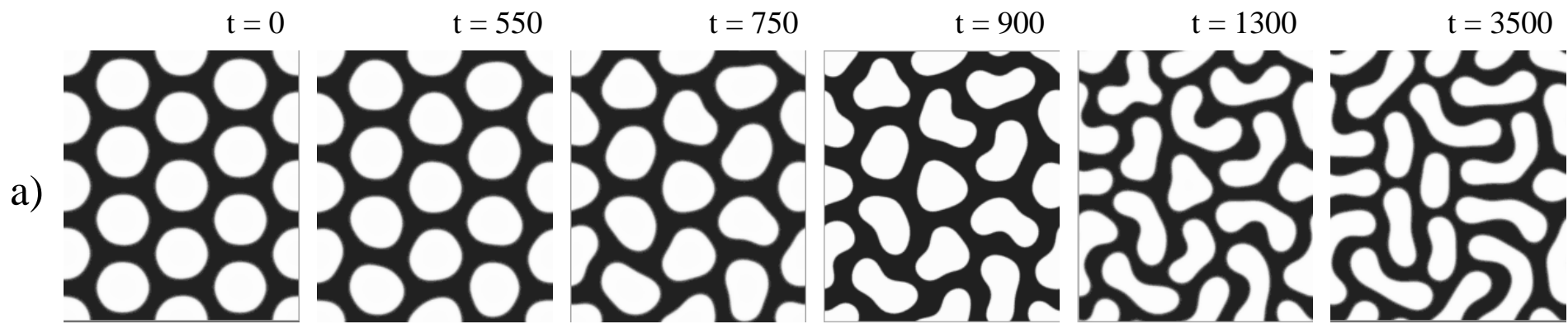
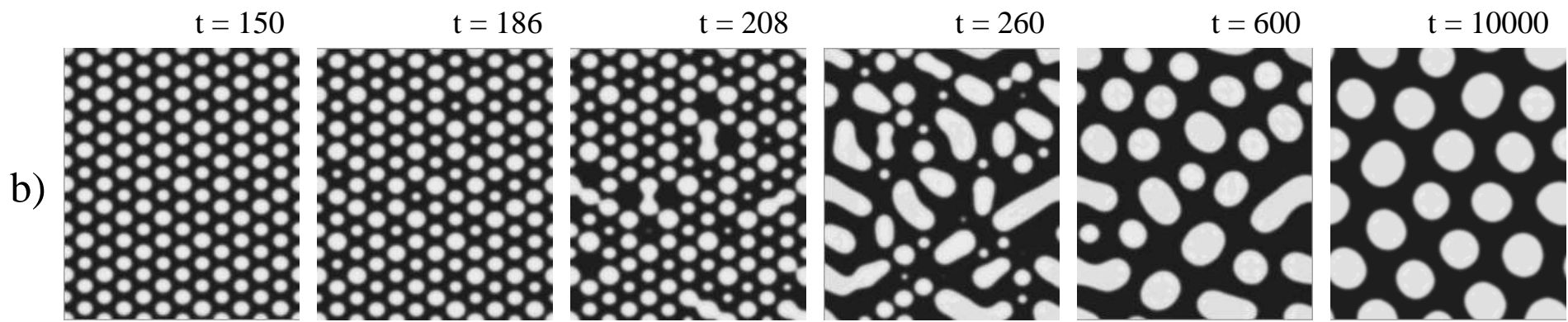
\end{document}